\documentclass[12pt,amsmath,amssymb]{revtex4}
\usepackage{amsmath,amsfonts,amsthm,amscd,amssymb,latexsym}
\usepackage{bm}
\usepackage{dcolumn}
\usepackage{graphicx}
\usepackage{epstopdf}
\usepackage{color}
\usepackage{epsf}
\usepackage{epsfig}
\usepackage{graphicx, epic, eepic, color}
\usepackage[colorlinks=true,urlcolor=blue,linkcolor=blue,citecolor=blue]{hyperref}

\begin{document}

\title{Generic Peculiar Motions in FLRW Spacetimes}

\author{Bahram \surname{Mashhoon}$^{1,2}$} 
\email{mashhoonb@missouri.edu}

\affiliation{
$^1$School of Astronomy, Institute for Research in Fundamental Sciences (IPM), Tehran 19395-5531, Iran\\
$^2$Department of Physics and Astronomy, University of Missouri, Columbia,
Missouri 65211, USA\\
}

\date{\today}
\begin{abstract}
In the standard Friedmann-Lema\^itre-Robertson-Walker (FLRW) spacetime, we consider a local cosmic test mass that is boosted in some direction relative to the standard comoving observers. The geodesic (Fermi) normal coordinate system established around the world line of the boosted cosmic mass is constructed within an approximation scheme and the resulting spacetime metric is compared with the corresponding metric of the Fermi system established around the world line of a comoving observer. The circular gravitomagnetic field around the direction of motion of the boosted cosmic mass is studied. 
\end{abstract}

\maketitle


\section{Introduction}

In the standard FLRW cosmology, peculiar motions refer to deviations from the Hubble flow caused by the attraction of gravity due to the presence of inhomogeneities. Therefore, peculiar motions are routinely treated  within the framework of cosmological perturbation theory~\cite{Baumann:2022mni, Mohayaee:2020wxf, TPA}. 

A different approach is adopted in the present work: We consider a cosmic test mass that is boosted with speed $c\,\beta(t)$ in some direction with respect to the Hubble flow and study the gravitational field in a quasi-inertial Fermi normal coordinate system established along the world line of the boosted cosmic mass, which then permanently occupies the spatial origin of the Fermi coordinates. The invariantly defined quasi-inertial Fermi system provides a proper approach for the description of local measurements within a spatial domain with distances small compared to the Hubble radius. Observational data indicate that in practice $\beta$ is small compared to unity; for instance,  the velocity dispersion $c\,\beta_{\rm pec}$ in clusters of galaxies indicates that 
$\beta_{\rm pec}$ is of order $10^{-3}$. For the study of Fermi coordinates within the context of standard cosmological models,  see~\cite{Mashhoon, Cooperstock:1998ny, Mashhoon:2007qm, KlCo, Pajer:2013ana, Dai:2015rda} and the references cited therein.  For recent related investigations, see~\cite{Tabatabaei:2024juy, Molaei:2024zxx}. 

A gravitational field is in general characterized via a spacetime metric, 
\begin{equation}\label{I1}
ds^2 = g_{\mu \nu}(x)\, dx^\mu \,dx^\nu\,, 
\end{equation}
where an event is denoted by admissible spacetime coordinates $x^\mu = (ct, x^i)$. The metric tensor $g_{\mu \nu}$ satisfies Einstein's gravitational field equations~\cite{Einstein}
\begin{equation}\label{I2}
G_{\mu \nu} + \Lambda\, g_{\mu \nu}=\kappa\,T_{\mu \nu}\,, 
 \end{equation}
where $T_{\mu \nu}$ is the symmetric energy-momentum tensor of matter, $\kappa:=8 \pi G/c^4$, $\Lambda$ is the cosmological constant and $G_{\mu \nu}$ is the Einstein tensor~\cite{R1, Griffiths:2009dfa}, 
\begin{equation}\label{I3}
G_{\mu \nu} := R_{\mu \nu}-\frac{1}{2} g_{\mu \nu}\,R\,, \qquad R = g^{\alpha \beta}R_{\alpha \beta}\,.
 \end{equation} 
Here, Greek indices run from 0 to 3, while Latin indices run from 1 to 3; moreover, the signature of the metric is +2 and we use natural units such that $c = G = 1$, unless specified otherwise. 

The Riemann curvature tensor can be decomposed into a source part related to Ricci and scalar curvatures and a part that is independent of the gravitational source; that is, 
\begin{equation}\label{I4}
R_{\mu \nu \rho \sigma} = C_{\mu \nu \rho \sigma} +g_{\mu[\rho}\,R_{\sigma]\nu} - g_{\nu[\rho}\,R_{\sigma]\mu} - \frac{1}{6}\,(g_{\mu \rho}\,g_{\nu \sigma} - g_{\mu \sigma}\,g_{\nu \rho})\,R\,,
\end{equation}
where $C_{\mu \nu \rho \sigma}$ is the  traceless Weyl conformal curvature tensor that represents the ``free" gravitational field.  The Riemann tensor satisfies the Bianchi identity $R_{\mu \nu[\rho \sigma; \delta]} = 0$.  At any point on the spacetime manifold, the Riemann tensor has in general 20 independent components, whereas the Ricci tensor has 10 independent components. Hence, the Weyl tensor has 10 independent components as well. 

We are interested in the standard Friedmann-Lema\^itre-Robertson-Walker (FLRW) cosmological models that are spatially homogeneous and isotropic. These spacetimes are conformally flat; hence, $C_{\mu \nu \rho \sigma} = 0$. 
We assume that the FLRW spacetime is given by
\begin{equation}\label{I5}
 ds^2 = - dt^2 + \frac{a^2(t)}{f^2(r)}\,\delta_{ij}\,dx^i\,dx^j\,, 
\end{equation} 
where $a(t)$ is the scale factor, $r = (\delta_{ij} x^i x^j)^{1/2}$ is the radial coordinate and   
\begin{equation}\label{I6}
f(r) = 1 + \frac{1}{4} k\, r^2\,.
\end{equation}
Here, $k = 1$, $-1$, or $0$, for the spatially closed, open, or flat model, respectively. The cosmic time $t$ is such that $ct$ and the scale factor $a(t)$ have dimensions of length, while the spatial coordinates $(x^1, x^2, x^3)$ are dimensionless. The connection coefficients for metric~\eqref{I5} are given in Appendix A and the construction of Fermi coordinates is briefly discussed in Appendix B. 

The Hubble parameter $H$ and the deceleration parameter $q$ are given by
\begin{equation}\label{I7}
H := \frac{1}{a} \frac{da}{dt} =\frac{\dot{a}}{a}\,, \qquad q H^2 = -\frac{\ddot{a}}{a} = -\dot{H} -H^2\,.
\end{equation}
It is usually assumed that the cosmic matter can be approximated by a perfect fluid of the form   
\begin{equation}\label{I8}
T^{\mu \nu} =  (\mu + p ) u^\mu u^\nu + p g^{\mu \nu}\,, 
\end{equation}
where $\mu$ is the energy density, $p$ is the pressure and $u^\mu = dx^\mu /dt$  is the unit 4-velocity vector of the perfect fluid that is comoving, namely, $u^\mu = \delta^{\mu}_{0}$.  
 Moreover, $T^{\mu \nu}{}_{; \nu} = 0$ implies 
\begin{equation}\label{I9}
(\mu + p ) u^{\nu}{}_{; \nu}  = -\frac{d\mu}{dt}\,  
\end{equation}
and
\begin{equation}\label{I10}
(\mu + p ) \frac{Du^\mu}{dt} = -(g^{\mu \nu} + u^\mu u^\nu) \frac{\partial p}{\partial x^\nu}\,. 
\end{equation}
Henceforth, we assume that $\mu + p  \ne 0$, $\mu = \mu (t)$, $p = p(t)$ and the Hubble flow is geodesic. De Sitter and anti-de Sitter spacetimes, where $\mu = p = 0$,  are considered in Appendices C and D, respectively. A detailed discussion of these spacetimes of positive (de Sitter) and negative (anti-de Sitter) constant curvatures is contained in~\cite{HE}. 

Let us briefly digress here and mention that $\mu + p$ in the present context has the interpretation of the effective density of mass-energy. To illustrate this concept in a dynamic situation, consider again the energy-momentum tensor $\mathfrak{T}^{\mu \nu}$ of a perfect fluid in Minkowski spacetime,
\begin{equation}\label{I10a}
\mathfrak{T}^{\mu \nu} =  (\mu' + p' ) \mathfrak{u}^\mu \mathfrak{u}^\nu + p' \eta^{\mu \nu}\,,   
\end{equation}
where
\begin{equation}\label{I10b}
\mathfrak{u}^\mu = \gamma' \,\left(1, \frac{\bm{\mathfrak{v}}}{c} \right)\,, \qquad \gamma' = \left[1 - \left(\frac{\mathfrak{v}}{c}\right)^2\right]^{-\frac{1}{2}}\,. 
\end{equation}
With 
\begin{equation}\label{I10c}
\varrho' := \frac{1}{c^2} \gamma'^2 (\mu' + p' )\,,  
\end{equation}
the conservation law $\mathfrak{T}^{\mu \nu}{}_{, \nu} = 0$ implies
\begin{equation}\label{I10d}
\frac{\partial \varrho'}{\partial t} + \bm{\nabla} \cdot (\varrho' \bm{\mathfrak{v}}) = \frac{1}{c^2}\frac{\partial p'}{\partial t}\,, \qquad \varrho' \, \frac{d\bm{\mathfrak{v}}}{dt} = - \bm{\nabla} p' - \frac{1}{c^2}\frac{\partial p'}{\partial t}\,\bm{\mathfrak{v}}\,,
\end{equation}
which are reminiscent of the continuity and Euler equations of Newtonian hydrodynamics, respectively. In a gravitational field, the spacetime is locally Minkowskian; hence, $\varrho'$ can be physically interpreted as the effective density of matter. We will use this concept later in Section III.

The gravitational field equations for the FLRW models imply
 \begin{equation}\label{I11}
U(t) := -\frac{\ddot{a}}{a} =  - \frac{1}{3} \,\Lambda + \frac{4 \pi}{3}\, (\mu +3\, p )\,,
\end{equation}
\begin{equation}\label{I12}
W(t) := H^2 + \frac{k}{a^2} = \frac{1}{3}\, \Lambda + \frac{8 \pi}{3} \,\mu\,. 
\end{equation}
These equations are consistent with Eq.~\eqref{I9}; indeed, we find $u^{\nu}{}_{; \nu} = \Gamma^\mu_{\mu 0} = 3 H$. 

The preferred comoving observers have a natural adapted orthonormal tetrad frame $\chi^{\mu}{}_{\hat \alpha}$ given by
\begin{equation}\label{I13}
\chi^{\mu}{}_{\hat  0} = \delta^\mu_0\,, \qquad \chi^{\mu}{}_{\hat i} = \frac{f(r)}{a(t)}\, \delta^\mu _i\,. 
\end{equation}
That is, $g_{\mu \nu}\,\chi^{\mu}{}_{\hat \alpha}\,\chi^{\nu}{}_{\hat \beta} = \eta_{\hat \alpha \hat \beta}$, where $\eta_{\hat \alpha \hat \beta}$ is the Minkowski metric given by diag$(-1, 1, 1, 1)$ in our convention.
It is straightforward to show, using the connection coefficients in Appendix A,  that this tetrad frame field is parallel transported along the geodesic world lines of the comoving observers. 

The curvature tensor as measured by a comoving observer is given by
\begin{equation}\label{I14}
R_{\hat \alpha \hat \beta \hat \gamma \hat \delta} = R_{\mu \nu \rho \sigma}\,\chi^{\mu}{}_{\hat \alpha}\,
\chi^{\nu}{}_{\hat \beta}\,\chi^{\rho}{}_{\hat \gamma}\,\chi^{\sigma}{}_{\hat \delta}\,.
\end{equation}
Taking advantage of the symmetries of the Riemann tensor, Eq.~\eqref{I14} can be represented by a $6\times6$ matrix $\mathcal{R} = (\mathcal{R}_{MN})$, where the indices $M$ and $N$ range over the set $(01,02,03,23, 31,12)$. Thus we can write
\begin{equation}
\label{I15}
\mathcal{R}=\left[
\begin{array}{cc}
\mathcal{E} & \mathcal{B}\cr
\mathcal{B^{\dagger}} & \mathcal{S}\cr
\end{array}
\right]\,,
\end{equation}
where $\mathcal{E}$  and $\mathcal{S}$ are symmetric $3\times3$ matrices and  $\mathcal{B}$ is traceless. The tidal matrix $\mathcal{E}$ represents the ``electric" components of the curvature tensor as measured by the fiducial observer, whereas $\mathcal{B}$ and $\mathcal{S}$ represent its ``magnetic" and ``spatial" components, respectively.  In Ricci-flat regions of spacetime, Eq.~\eqref{I15} simplifies, since   
$\mathcal{S} = -\mathcal{E}$, $\mathcal{E}$ is traceless and $\mathcal{B}$ is symmetric. Hence, the Weyl curvature tensor is completely determined by the 10 independent components of the symmetric and traceless $3\times3$ matrices that are its ``electric" and ``magnetic" components. 

For the standard FLRW models, it is possible to show that $\mathcal{R}$ is diagonal such that 
\begin{equation}\label{I16}
\mathcal{E} = U\,\mathcal{I}\,, \qquad \mathcal{B} = 0\,, \qquad \mathcal{S} = W\,\mathcal{I}\,.
\end{equation}
Here, $U$ and $W$ are given by the field equations~\eqref{I11} and~\eqref{I12}, respectively, and $\mathcal{I}$ is the unit $3\times3$ matrix. 

A comoving observer can employ an invariantly defined quasi-inertial geodesic coordinate system it its neighborhood for local measurements. Such a Fermi normal coordinate system is discussed in Appendix B. Here, it is convenient to express Cartesian Fermi coordinates $X^{\hat \mu}$ as 
\begin{equation}\label{I17}
X^{\hat \mu} = (T, X, Y, Z)\,.
\end{equation}
In these coordinates,  the reference observer is permanently fixed at the spatial origin of the Fermi coordinate system and thus has Fermi coordinates $(T, 0, 0, 0)$, where $T = t$ is the cosmic time in this case. Using the results of Appendix B, we now  choose a specific comoving observer and establish a Fermi normal coordinate system $(T, X, Y, Z)$ along its world line. The resulting approximate Fermi metric is valid to second order in the spatial Fermi coordinates and is given by
\begin{align}\label{I18}
\nonumber   ^Fds^2_{\rm comoving~observer} = &{} - [1+ U(T)\,\rho^2] \, dT^2 + dX^2 + dY^2 + dZ^2   \\
& {} -\frac{1}{3} W(T) \left[(X\,dY-Y\,dX)^2 + (X\,dZ-Z\,dX)^2 + (Y\,dZ-Z\,dY)^2\right]\,,
\end{align}
where $\rho$,
\begin{equation}\label{I19}
\rho := (X^2 + Y^2 + Z^2)^{\frac{1}{2}}\,
\end{equation}
is the Fermi radial coordinate. The space in metric~\eqref{I18} is clearly isotropic; therefore, it is convenient to introduce a spherical polar coordinate system $(\rho, \vartheta, \varphi)$ such that, for instance, 
\begin{equation}\label{I20}
X = \rho \,\cos \vartheta\,, \qquad Y = \rho\, \sin \vartheta \,\cos \varphi \,, \qquad Z = \rho\, \sin \vartheta \,\sin \varphi\,.
\end{equation}
Then, we find the spherically symmetric Fermi metric
\begin{align}\label{I21}
^Fds^2_{\rm comoving\,observer} = - [1+ U(T)\,\rho^2]\, dT^2 + d\rho^2 + \rho^2\, \left[1-\frac{1}{3} W(T)\,\rho^2\right]\,d\Omega^2\,,
\end{align}
where
\begin{equation}\label{I22}
d\Omega^2 = d\vartheta^2 + \sin^2 \vartheta \, d\varphi^2\,.
\end{equation}

We are interested in using these results in connection with certain peculiar motions. In previous work, cosmological Fermi systems were mainly employed in connection with estimating the physical influence of the recession of faraway galaxies on local systems~\cite{Mashhoon, Cooperstock:1998ny, Mashhoon:2007qm}. 


\subsection{Directional Outflows Relative to Comoving Observers}

For a physical application of these ideas, imagine a galaxy that approximately follows the Hubble flow. The galaxy's core contains a central engine involving a supermassive Kerr black hole with an accretion disk and prominent bipolar jets as in active galactic nuclei (AGNs). The tidal influence of the Kerr source on the motion of particles within the jet relative to the central engine have been investigated in detail in~\cite{BCM} and the references therein. In these studies, the focus has been on purely gravitational effects and important electromagnetic forces have been ignored for the sake of simplicity. In the present work, we are interested in the gravitational influence of the rest of the FLRW universe on the motion of the jet relative to the comoving galaxy. In this case, we employ Fermi metric~\eqref{I18} and write the geodesic equation of motion of a free particle relative to the fiducial comoving observer that is fixed at the spatial origin of the Ferm system. To first order in the spatial Fermi coordinates, the general equation of motion is given by Eq.~\eqref{B16} of Appendix B; furthermore, for motion that is restricted to a definite direction, Eq.~\eqref{B16} reduces to Eq.~\eqref{B17}. Based on this result, the equation of motion in the special case of a bipolar jet is independent of spatial direction due to the spatial isotropy of the background and is given by
\begin{equation}\label{G1}
\ddot{\Psi} + qH^2 (1- 2\,\dot{\Psi}^2) \Psi = 0\,, 
\end{equation} 
where $qH^2 = -\ddot{a}/{a} = U(t)$ and we have ignored higher-order terms in the spatial Fermi coordinate $\Psi$. Moreover, $T = t$,  $\dot{\Psi} := d\Psi/dt$, etc. The general behavior of the solutions of this nonlinear ordinary differential equation has been discussed in~\cite{Chicone:2002kb, Mashhoon:2020tha}. Let us note the existence  of an exact solution involving uniform rectilinear motion at the constant critical speed $v_{c} = c/\sqrt{2} \approx 0.707 c$; that is,  
\begin{equation}\label{G2}
\Psi (t) = \Psi (t_i) \pm\, (2)^{-\frac{1}{2}} c\,(t-t_i)\,, 
\end{equation}
where $t_i$ is an initial instant of cosmic time. In addition, Eq.~\eqref{G1} is invariant under $\Psi \to -\Psi$, a feature that is consistent with the existence of bipolar jets. Finally, there is a rest point at $(\Psi, \dot{\Psi}) = (0, 0)$; that is, there is no motion once the initial speed is zero at $\Psi = 0$. In the case under discussion, Fermi system is admissible for $|\Psi| \ll L_{H}$, where $L_{H} =c/H$ is the Hubble radius. 

Imagine free jet motion radially away from the fiducial observer. In the current standard cosmological model,  the deceleration parameter at the present epoch $t = t_0$ is such that  $q_0 \approx -0.55$. For negative $q$,  Eq.~\eqref{G1} implies that the character of the motion is toward the exact solution of speed $v_c$ as the local attractor~\cite{Chicone:2002kb, Mashhoon:2020tha}. This means that if the initial jet speed is less (greater) than $v_c$, the jet accelerates (decelerates) toward $v_c$. For positive $q$, however, the character of the motion is away from the local attractor. We must emphasize the approximate nature of these results, as we have neglected higher-order terms in the  spatial Fermi distance.  The influence of the higher-order tidal terms has been investigated in certain simple cases by the construction of exact Fermi coordinate systems~\cite{Chicone:2005vn}. Exact Fermi coordinate systems are discussed in Appendices C and D for de Sitter and anti-de Sitter spacetimes, respectively. 


\section{Boosted Cosmic Mass}

Let us now imagine that a cosmic mass is boosted with speed $c\,\beta(t)$ relative to the background comoving observers in a given direction. We take this direction to be along the $x$ axis with no loss in generality due to the spatial isotropy of the background FLRW spacetime. The boosted cosmic mass has an adapted orthonormal tetrad frame $e^{\mu}{}_{\hat \alpha}$ given by
\begin{equation}\label{M1}
e^{\mu}{}_{\hat 0} = \gamma ( \chi^{\mu}{}_{\hat 0} + \beta\, \chi^{\mu}{}_{\hat 1})\,,
\end{equation}
\begin{equation}\label{M2}
e^{\mu}{}_{\hat 1} = \gamma ( \chi^{\mu}{}_{\hat 1} + \beta\, \chi^{\mu}{}_{\hat 0})\,,
\end{equation}
\begin{equation}\label{M3}
e^{\mu}{}_{\hat 2} =  \chi^{\mu}{}_{\hat 2}\,, \qquad e^{\mu}{}_{\hat 3} = \chi^{\mu}{}_{\hat 3}\,.
\end{equation}
Here, 
\begin{equation}\label{M4}
\gamma = (1-\beta^2)^{-\frac{1}{2}} = \frac{dt}{d\eta}\,
\end{equation}  
is the Lorentz factor and
\begin{equation}\label{M5}
e^{\mu}{}_{\hat 0} = \bar{u}^\mu = \frac{d\bar{x}^\mu}{d\eta}\,,
\end{equation}
where $\eta$ is the proper time of the boosted cosmic mass. In $(t, x^i)$ coordinates, we have
\begin{equation}\label{M6}
e^{\mu}{}_{\hat 0} = \left[\gamma, \gamma\,\beta\,\frac{f(r)}{a(t)}, 0, 0\right]\,.
\end{equation}

Let $\bar{\mathcal{R}}$ be the Riemann curvature tensor as measured by the boosted observer~\cite{BM1, BM2}; that is, 
\begin{equation}\label{M7}
\bar{R}_{\hat \alpha \hat \beta \hat \gamma \hat \delta} = R_{\mu \nu \rho \sigma}\,e^{\mu}{}_{\hat \alpha}\,e^{\nu}{}_{\hat \beta}\,e^{\rho}{}_{\hat \gamma}\,e^{\sigma}{}_{\hat \delta}\,.
\end{equation} 
In general, the behavior of the measured components of the Riemann curvature tensor under boosts can be determined based on the results given in Ref.~\cite{BM2}. In the present case, direct calculation of $\bar{\mathcal{R}}$ is straightforward. The result is a symmetric $6\times6$ matrix that can be expressed in terms of $\bar{\mathcal{E}}$, $\bar{\mathcal{B}}$ and $\bar{\mathcal{S}}$ as follows: 
\begin{equation}\label{M8}
\bar{\mathcal{E}} = {\rm diag} (U, \mathcal{U}, \mathcal{U})\,,  \qquad \mathcal{U} := \gamma^2 (U + \beta^2 \,W)\,,
\end{equation}
\begin{equation}\label{M9}
\bar{\mathcal{B}} = \gamma^2\,\beta (U + W)\,\mathcal{J}_1\,,
\end{equation}
where $\mathcal{J}_1$ is given by
\begin{equation}\label{M10}
\mathcal{J}_1 = \left[
\begin{array}{ccc}
0 & 0 & 0 \cr
0 & 0 & 1 \cr
0 &-1 & 0 \cr
\end{array}
\right]\,,
\end{equation} 
and
\begin{equation}\label{M11}
\bar{\mathcal{S}} = {\rm diag} (W, \mathcal{W}, \mathcal{W})\,,  \qquad \mathcal{W} := \gamma^2 (W + \beta^2 \,U)\,.
\end{equation}
Let us note that $\mathcal{J}_i $ is a $3\times3$ matrix given by $(\mathcal{J}_i)_{jk} = \epsilon_{ijk}$, which is proportional to the operator of infinitesimal rotations about the $x^i$ axis. 

The curvature tensor of the FLRW universe as determined by the boosted observer illustrates certain general features of the boosted gravitational field. In general, it turns out that the component of the field along the direction of the boost is not affected, while in the transverse directions the field is enhanced by a factor of $\gamma^2$. More explicitly,  under the boost the elements of $\mathcal{E}$, $\mathcal{B}$ and $\mathcal{S}$ in the direction parallel to the direction of the boost are left unchanged, whereas those perpendicular to the direction of the boost are enhanced by $\gamma^2$; moreover, the mixed elements are enhanced by a factor of $\gamma$~\cite{BM1, BM2}. This circumstance is reminiscent of the behavior of the (spin-1) electromagnetic field under a boost: The components of the electric field ($\mathbb{E}$) and magnetic field ($\mathbb{B}$) parallel to the direction of the boost remain the same as before, while those perpendicular to the direction of the boost are enhanced by a factor of $\gamma$. Under the boost, the field strength of a field of spin-s goes in general as  $\gamma^s$; however, there are exceptional circumstances that are briefly mentioned below. The strength of the gravitational field can be augmented in general by a factor of $\gamma^2$ under a boost due to the spin-2 nature of the gravitational interaction; alternatively, one can say that the radius of curvature of spacetime measured by the boosted observer is Lorentz contracted~\cite{BM1, BM2}, except in certain special cases. It follows that a gravity gradiometer would in general measure extremely strong tidal forces when it moves very fast with $\beta \to 1$ and $\gamma \gg 1$; however, tidal forces remain finite along certain spatial directions such as the radial direction in the exterior Schwarzschild spacetime~\cite{BM1, BM2}.  Along such a \emph{special tidal direction}, the corresponding world line of the boosted observer as $\beta \to 1$ approaches a null direction. The special tidal directions are thus associated with certain tidally \emph{nondestructive null directions} in spacetime. The basic mathematical connection between the special tidal directions and the principal null directions of the curvature tensor has been established by Beem and Parker~\cite{BePa} and Hall and Hossack~\cite{HaHo}. Our explicit results~\eqref{M8}--\eqref{M11} indicate that there are no special tidal directions in any of the standard FLRW cosmological models. On the other hand, every direction is a special tidal direction in a spacetime of constant nonzero curvature, namely, de Sitter ($\Lambda > 0$) or anti-de Sitter ($\Lambda < 0$) universe, where
\begin{equation}
\label{M12}
\mathcal{R}= - \frac{1}{3} \, \Lambda\,\left[
\begin{array}{cc}
\mathcal{I} & 0 \cr
0 & - \mathcal{I} \cr
\end{array}
\right]\,.
\end{equation}

It is interesting to note that a cosmic mass boosted with $\gamma \gg 1$ in an arbitrary direction is generally subject to strong tidal forces $\propto \gamma^2$ in directions perpendicular to the direction of motion in accordance with the Jacobi equation. Under appropriate circumstances, the tidal forces for $\gamma \gg 1$ tend to tear the cosmic matter apart in the transverse directions resulting in a highly collimated outflow.

We can establish an approximate Fermi normal coordinate system along the accelerated world line of the boosted cosmic mass. The orthonormal adapted tetrad frame moves along the world line in accordance with 
\begin{equation}\label{M13}
\frac{D e^{\mu}{}_{\hat \alpha}}{d\eta} = \Phi_{\hat \alpha}{}^{\hat \beta}\, e^{\mu}{}_{\hat \beta}\,, 
\end{equation}
where $\Phi_{\hat \alpha \hat \beta} = - \Phi_{\hat \beta \hat \alpha}$ is the antisymmetric acceleration tensor. Its nonzero components in this case are given by $\gamma^2 d\beta/d\eta$; that is, 
\begin{equation}\label{M14}
\Phi_{\hat 0 \hat 1}  = -\Phi_{\hat 1 \hat 0} = \gamma^3 \frac{d\beta}{dt}\,, 
\end{equation}
using the results of Appendix A. Indeed, the projection of the translational acceleration of the boosted cosmic mass upon its spatial frame can be expressed as 
\begin{equation}\label{M15}
\frac{D e^{\mu}{}_{ \hat 0}}{d \eta }\, e_{\mu \,\hat i} = \Phi_{\hat 0 \hat i}  = (\gamma^3 \frac{d\beta}{dt}, 0, 0)\, 
\end{equation}
and the proper time along the path $\eta$ is given by $d\eta/dt = 1/\gamma$; that is, 
\begin{equation}\label{M16}
\eta  = \int^t \left[1-\beta^2(\zeta)\right]^{\frac{1}{2}} d\zeta\,. 
\end{equation}

Next, using the results of Appendix B, we find the metric of the Fermi coordinate system to second order in the spatial Fermi distance is given by
\begin{align}\label{M17}
\nonumber   ^Fds^2_{\rm boosted\,observer} = &{} - \left\{\left[1+ \gamma^2\,\frac{d\beta}{dT}\,X \right]^2 + U(T) X^2 + \mathcal{U}(T)\,(Y^2+Z^2)\right\}\,dT^2   \\
\nonumber &{} - \frac{4}{3} \mathcal{V}(T)\,[(Y^2+Z^2)dX -X(YdY + ZdZ)]\,dT + dX^2 + dY^2 + dZ^2  \\
\nonumber & {} -\frac{1}{3}\,\mathcal{W}\,\left[(X\,dY-Y\,dX)^2 + (X\,dZ-Z\,dX)^2\right]   \\
& {}-\frac{1}{3}\, W(T)\,(Y\,dZ-Z\,dY)^2\,,
\end{align}
where $T = \eta$ and
\begin{align}\label{M18}
\mathcal{V} := \gamma^2\, \beta (U + W)\,, \qquad   \mathcal{U} - U = \mathcal{W} - W = \beta\,\mathcal{V}\,.
\end{align}
The Fermi metric~\eqref{M17} is valid for a constant boost $\beta$ as well, except that the acceleration term in the metric proportional to $d\beta/dT$ vanishes in this case.  

The boosted cosmic mass resides at the spatial origin of the Fermi coordinate system and the spacetime is Minkowskian in its immediate neighborhood. The Fermi metric expresses the deviation of spacetime from the Minkowski metric to second order in the spatial distance away from the boosted cosmic mass. The approximate nature of the Fermi metric makes it possible to compare it with the gravitoelectromagnetic (GEM) metric of  the general linear approximation of general relativity~\cite{Mash93, Mashhoon:2003ax, Mashhoon:2000he}. That is,  
\begin{equation}\label{M19}
^{GEM}ds^2 = - (1+ 2 \mathbb{V}_g)\, dT^2 -4 (\delta_{\hat i \hat j} \mathbb{A}^{\hat i}_g\, dX^{\hat j})\,dT + (1-2 \mathbb{V}_g)\,\delta_{\hat i \hat j} dX^{\hat i} dX^{\hat j}\,.
\end{equation} 
 Thus, ignoring the spatial parts, the resulting gravitoelectric and gravitomagnetic potentials of the Fermi metric are given by
\begin{equation}\label{M20}
 \mathbb{V}_g = \gamma^2\,\frac{d\beta}{dT}\,X + \frac{1}{2}\,\left(\gamma^2\,\frac{d\beta}{dT}\right)^2 X^2 + \frac{1}{2} U X^2 + \frac{1}{2} \mathcal{U} \,(Y^2+Z^2)\, 
\end{equation} 
and
\begin{equation}\label{M21}
\mathbb{A}_g = \frac{1}{3}\,c^2 \gamma^2 \beta\,\,(U+W)\,(Y^2 + Z^2, -XY, -XZ)\,,
\end{equation} 
respectively. Let us note that 
\begin{equation}\label{M21a}
U+W = \frac{4 \pi G}{c^4} (\mu + p)\,.
\end{equation}

Observational evidence regarding peculiar motions indicates that  in practice $\beta \ll 1$; therefore, we focus on terms linear in $\beta$ and ignore higher-order terms. It follows from $T = \eta$ in this case and Eq.~\eqref{M16} that 
\begin{equation}\label{M22}
T = t + \mathcal{O}(\beta^2)\,.
\end{equation}
Similarly, we have $\mathcal{U} = U + \mathcal{O}(\beta^2)$ and $\mathcal{W} = W + \mathcal{O}(\beta^2)$, so that the spatial Fermi metric turns out to be the same as that of the comoving observer when we ignore $\beta^2$ terms. This means that deviations from spatial isotropy occur due to the presence of terms of  order $\beta^2$; in any case, the azimuthal symmetry about the direction of motion of the boosted cosmic mass persists. Moreover, the Jacobi equation in this case is given by
\begin{equation}\label{M23}
\frac{d^2 X^{\hat i}}{dt^2} =  -\partial_{\,\hat i} \mathbb{V}_g\,,
\end{equation}
so that we have explicitly,
\begin{equation}\label{M24}
\frac{d^2X}{dt^2} + \frac{d\beta}{dt} \left( 1 + \frac{d\beta}{dt}\,X\right) + U X = 0\,, \quad \frac{d^2Y}{dt^2} + U Y = 0\,, \quad \frac{d^2Z}{dt^2} + U Z = 0\,,
\end{equation}  
where we have assumed that $d\beta/dt$ may not be small.  These equations describe the motion of free nearby test particles relative to the boosted cosmic mass. Along the direction of the boost, the equation for $X$ exhibits inertial forces that come about due to the accelerated nature of the Fermi frame, while along the transverse directions the equations of motion are not affected by the boost to first order in $\beta$. Let us consider the differential equations for $Y$ and $Z$ in Eq.~\eqref{M24}, where $U(t) = - \ddot{a}/a$, and let $\Upsilon(t)$ be a solution of such an equation. Then, 
\begin{align}\label{M25}
\Upsilon(t) = a(t) \left\{\Upsilon(t_0) + [\dot{\Upsilon}(t_0)- H_0\,\Upsilon(t_0)]\,\int_{t_0}^t\frac{d\zeta}{a^2(\zeta)}\right\}\,,
\end{align}
where  $t = t_0$ is the present epoch such that $a(t_0) = 1$ and $H_0 = H(t_0)$ is the corresponding Hubble parameter. For instance, in the case of the Einstein-de Sitter universe,  $a(t) = (t/t_0)^{2/3}$ and we find
\begin{align}\label{M26}
\Upsilon(t)  = [2\Upsilon(t_0) -3t_0 \dot{\Upsilon}(t_0)] \left(\frac{t}{t_0}\right)^{1/3} - [\Upsilon(t_0) -3t_0 \dot{\Upsilon}(t_0)] \left(\frac{t}{t_0}\right)^{2/3}\,.
\end{align}

Finally, it is useful to write metric~\eqref{M17} to first order in $\beta$ in terms of spherical Fermi coordinates using Eq.~\eqref{I20}; with $\dot{\beta} := d\beta/dT$, we find
\begin{align}\label{M27}
\nonumber   ^Fds^2_{\rm boosted\,observer} \approx &{} - \left[(1+\dot{\beta}\,\rho \cos\vartheta)^2 + U(T) \rho^2 \right] dT^2 + \frac{4}{3} \beta\,(U+W)\rho^3\, \sin\vartheta \,d\vartheta\, dT  \\
& {} + d\rho^2+ \rho^2\, \left[1-\frac{1}{3} W(T)\,\rho^2\right]\,d\Omega^2\,,
\end{align}
which should be compared with Eq.~\eqref{I21}.


\section{Gravitomagnetic Field of the Boosted Cosmic Mass}

The connection between electricity and magnetism was first discovered in 1820 by the Danish physicist Hans Christian {\O}rsted (1777-1851), who observed a circular magnetic presence around a wire that carried electric current. This discovery led to further work by the French physicists J.-P. Biot, F. Savart and A.-M. Amp\`ere~\cite{Whittaker}.  

The circular magnetic field around a straight wire carrying constant electric current $I$ in the $x$ direction expressed in Gaussian units is given by 
\begin{equation}\label{N1}
\mathbb{B} = \frac{2\,I}{c\,r^2}\,(0, -z, y)\,, 
\end{equation}
where $r = (y^2+z^2)^{\frac{1}{2}}$ is the radius of the circle in the $(y, z)$ plane.  

The gravitomagnetic field, i.e. $\mathbb{B}_g = \bm{\nabla} \times \mathbb{A}_g$,  corresponding to the mass current caused by the generic peculiar motion with speed $\beta (t)$ in the $x$ direction is given in the corresponding Fermi frame by 
\begin{equation}\label{N2}
\mathbb{B}_g =  c^2\,\gamma^2 \,\beta\, (U+W)\, ( 0, Z, -Y) = \frac{4 \pi G}{c^2} \gamma^2\beta\,(\mu + p)\, (0, Z, -Y)\,.
\end{equation}

As discussed in Section I, let us define $\varrho$,
\begin{equation}\label{N3}
\varrho = \frac{1}{c^2} \gamma^2 \,(\mu + p)\,, 
\end{equation}
to be the effective density of cosmic matter. Then, 
\begin{equation}\label{N4}
\mathbb{B}_g = - 4 \pi G \varrho \,\beta\, (0, -Z, Y)\,,
\end{equation}
where $\varrho(t) \,\beta(t)$ is proportional to the matter current in analogy with the electric current in Eq.~\eqref{N1}.  

In our discussion of gravitomagnetism in the context of the standard FLRW cosmology, the boost $\beta$ can be constant or dependent upon cosmic time $t$. It is interesting to note that the gravitomagnetic field of the boosted cosmic mass is independent of the cosmological constant. The direction of $\mathbb{B}_g$ is opposite to that of $\mathbb{B}$, as expected~\cite{Mash93, Mashhoon:2003ax, Mashhoon:2000he}. Furthermore, it is clear that $\mathbb{B}_g/c$ is locally equivalent to the angular velocity of rotation in agreement with the gravitational Larmor theorem~\cite{Mash93}. In the approximate Fermi system under consideration here, the gravitomagnetic field vanishes at the position of the cosmic mass (located at the spatial origin of Fermi coordinate system) and its magnitude increases linearly with the radius of the circular loop away from the direction of the boosted cosmic mass. This radius is expected be rather small in comparison with the Hubble radius $L_H = c/H$.  

Earth's gravitomagnetic field has been measured via the Gravity Probe B (GP-B) space experiment~\cite{Francis1, Francis2}. For a recent discussion of the astrophysical aspects of gravitomagnetism, see~\cite{Mashhoon:2024wvp} and the references therein. Within the context of cosmology, the possibility of detecting gravitational lensing caused by large-scale gravitomagnetic fields has been the subject of recent investigations~\cite{BHLC, Murray:2025wtc, Beordo:2025cpw}.  


\section{DISCUSSION}

In a FLRW cosmological model, a cosmic test mass is boosted with speed $\beta(t)$ in some direction with respect to the Hubble flow. From the standpoint of the boosted observer, space is axially symmetric about the direction of the boost and the deviation from spatial isotropy is proportional to $\beta^2$. Moreover, there is a circular gravitomagnetic field $\mathbb{B}_g$ around the direction of the boosted cosmic mass in close analogy with electrodynamics. The magnitude of  $\mathbb{B}_g$ is proportional to the current of matter in the boosted cosmic mass and is independent of the presence of the cosmological constant. 



\appendix

\section{Christoffel symbols for metric~\eqref{I5}}

Let us consider the FLRW metric~\eqref{I5} in Cartesian coordinates $x^\mu = (t, x^i)$. The nonzero connection coefficients, i.e. $\Gamma^{\mu}_{\nu \rho} =  \Gamma^{\mu}_{\rho \nu}$,   can be calculated using
\begin{equation}\label{A1}
 \Gamma^{0}_{i j} = \frac{a\,\dot{a}}{f^2}\, \delta_{ij}\,, \qquad \Gamma^{i}_{0 j} = \Gamma^{i}_{j 0} = \frac{\dot{a}}{a}\,\delta^i_j\,
\end{equation} 
and 
\begin{equation}\label{A2}
\Gamma^{i}_{j l} = \frac{k}{2f}\, (x^i \,\delta_{j l}  - x^j\, \delta_{i l} - x^l\, \delta_{i j})\,.
\end{equation}

\section{Fermi Coordinates}

In a given gravitational field, we consider  a congruence of future-directed timelike world lines.  We then choose a fiducial path $\tilde{x}^\mu(\tau)$ in the congruence, where $\tau$ is the proper time of the reference observer.  The adapted orthonormal  tetrad frame $\lambda^{\mu}{}_{\hat \alpha} (\tau)$ of the reference observer is such that
\begin{equation}\label{B1}
g_{\mu \nu} \,\lambda^\mu{}_{\hat \alpha}\,\lambda^\nu{}_{\hat \beta}= \eta_{\hat \alpha \hat \beta}\,, \qquad \lambda^{\mu}{}_{\hat 0} (\tau) = \frac{d\tilde{x}^\mu}{d\tau}\,.
\end{equation}
The adapted tetrad is carried along $\tilde{x}^\mu$ in accordance with
\begin{equation}\label{B2}
\frac{D\lambda^{\mu}{}_{\hat \alpha}}{d\tau} = \phi_{\hat \alpha}{}^{\hat \beta} \,\lambda^{\mu}{}_{\hat \beta}\,,
\end{equation}
where $\phi_{\hat \alpha \hat \beta}$ is the antisymmetric acceleration tensor. In analogy with the electromagnetic field tensor, $\phi_{\hat \alpha \hat \beta} \to (-\bm{A}, \bm{\Omega})$, where the ``electric" and ``magnetic" parts are the translational and rotational accelerations of the fiducial observer, respectively. More explicitly, 
\begin{equation}\label{B3}
\frac{d^2\tilde{x}^{\mu}}{d\tau^2}+\Gamma^\mu_{\nu \sigma}\,\frac{d\tilde{x}^\nu}{d\tau}\,\frac{d\tilde{x}^\sigma}{d\tau} = \mathcal{A}^\mu\,, \quad \mathcal{A}^\mu = A^{\hat i}\,\lambda^{\mu}{}_{\hat i}\,
\end{equation}
and $(A^{\hat 1}, A^{\hat 2}, A^{\hat 3}) = \bm{A}$, while $\bm{\Omega}$ is the angular velocity of rotation of the spatial frame  of the fiducial observer with respect to a locally nonrotating frame that is Fermi-Walker transported along  $\tilde{x}^\mu(\tau)$. 

At each event $\mathbb{O}$ along the reference path, we consider all spacelike geodesics that start from $\mathbb{O}$ and are orthogonal to the fiducial world line. On the resulting local spacelike hypersurface, we choose an event $\mathbb{P}$ with spacetime coordinates $x^\mu$ that can be connected to $\mathbb{O}$ by a unique spacelike geodesic of proper length $\sigma$. Let $\sigma = 0$ denote event $\mathbb{O}$ at proper time $\tau$ and $\xi^\mu (\tau) = (dx^\mu/d\sigma)_{\sigma = 0}$ be the unit spacelike tangent vector at $\mathbb{O}$ such that $\xi_\mu \,\lambda^{\mu}{}_{\hat 0} = 0$. Event $\mathbb{P}$ with spacetime coordinates $x^\mu$ is assigned Fermi normal coordinates $X^{\hat \mu}$, where~\cite{Synge, mash77}
\begin{equation}\label{B4}
X^{\hat 0} := \tau\,, \qquad X^{\hat i} := \sigma\, \xi^\mu(\tau)\, \lambda_{\mu}{}^{\hat i}(\tau)\,.
\end{equation}
Fermi coordinate are admissible within a cylindrical spacetime region along the fiducial world line.  The radius of this cylinder is a certain radius of curvature of spacetime~\cite{Chicone:2002kb, Chicone:2005vn}.

Henceforward, we write $X^{\hat 0} := T$ and $(X^{\hat 1}, X^{\hat 2}, X^{\hat 3}) = \bm{X}$ for simplicity and express the spacetime metric in Fermi coordinates as 
\begin{equation} \label{B5}
 ^Fds^2= g_{\hat \mu \hat \nu}(T, \mathbf{X})\,dX^{\hat \mu} \,dX^{\hat \nu}\,. 
\end{equation}
The coordinate transformation that connects $x^\mu$ to $X^{\hat \mu}$ can be given exactly in certain special situations~\cite{Chicone:2005vn, KC}; in this connection, see Appendices C and D. In general, however, it is only possible to express the Fermi metric as a series in powers of the spatial Fermi coordinates. This expansion is given to second order in the present work. That is, 
\begin{eqnarray}\label{B6}
g_{\hat 0 \hat 0} &=& -P^2 + Q^2  - \,^FR_{\hat 0 \hat i \hat 0 \hat j}\,X^{\hat i}\,X^{\hat j} + \mathcal{O}(|\bm{X}|^3)\,,\\
g_{\hat 0 \hat i} &=& Q_{\hat i} -\frac{2}{3} \,^FR_{\hat 0 \hat j \hat i \hat k}\,X^{\hat j}\,X^{\hat k} + \mathcal{O}(|\bm{X}|^3)\,,\label{B7}\\
g_{\hat i \hat j} &=& \delta_{\hat i \hat j} -\frac{1}{3} \,^FR_{\hat i \hat k \hat j \hat l}\,X^{\hat k}\,X^{\hat l} + \mathcal{O}(|\bm{X}|^3)\,,\label{B8}
\end{eqnarray}
where  $P$ and $\mathbf{Q}$ have to do with the translational and rotational accelerations of the reference observer, respectively. Indeed, 
\begin{equation}\label{B9}
P := 1 + \bm{A}(T) \cdot \bm{X}\,, \qquad \bm{Q} := \bm{\Omega}(T) \times \bm{X}\,.
\end{equation}
Furthermore, 
\begin{equation}\label{B10}
^FR_{\hat \alpha \hat \beta \hat \gamma \hat \delta}(T) := R_{\mu \nu \rho \sigma}\,\lambda^{\mu}{}_{\hat \alpha}\,
\lambda^{\nu}{}_{\hat \beta}\,\lambda^{\rho}{}_{\hat \gamma}\,\lambda^{\sigma}{}_{\hat \delta}\,.
\end{equation}


\subsection{Reduced Geodesic Equation}

It is interesting to discuss timelike and null geodesics of spacetime within the Fermi system. The equation of motion of a free test particle is given by
\begin{equation}\label{B11}
\frac{d^2 X^{\hat \mu}}{d\theta^2}+\Gamma^{\hat \mu}{}_{\hat \alpha \hat \beta}\,\frac{dX^{\hat \alpha}}{d\theta}\,\frac{dX^{\hat \beta}}{d\theta}=0\,,
\end{equation}
where $\theta$ is its proper time. Moreover, the free test particle's 4-velocity vector can be written as
\begin{equation}\label{B12}
\frac{dX^{\hat \mu}}{d\theta} = \Gamma\,(1, \bm{V})\,, \qquad    \bm{V} = \frac{d \bm{X}}{dT}\,,
\end{equation}
where  $\Gamma$ is the Lorentz factor
\begin{equation}\label{B13}
\Gamma := \frac{dT}{d\theta} = (- g_{\hat 0 \hat 0}-2\, g_{\hat 0 \hat i}\,V^{\hat i} - g_{\hat i \hat j}\,V^{\hat i}\,V^{\hat j})^{-\frac{1}{2}}\,. 
\end{equation}
The temporal and spatial parts of Eq.~\eqref{B11} can be combined to get the reduced geodesic equation~\cite{Chicone:2002kb}
\begin{equation}\label{B14}
\frac{d^2 X^{\hat i}}{dT^2}+\left(\Gamma^{\hat i}{}_{\hat \alpha \hat \beta}-\Gamma^{\hat 0}{}_{\hat \alpha \hat \beta}V^{\hat i} \right) \frac{dX^{\hat \alpha}}{dT}\frac{dX^{\hat \beta}}{dT}=0\,.
\end{equation}
This result is valid for a null geodesic as well provided
\begin{equation}\label{B15}
 g_{\hat 0 \hat 0} +2\, g_{\hat 0 \hat i}\,V^{\hat i} + g_{\hat i \hat j}\,V^{\hat i}\,V^{\hat j} = 0\,. 
\end{equation}
The fiducial observer is at the spatial origin of the Fermi coordinate system and the spacetime is Minkowskian in its neighborhood; hence, the Fermi velocity $\bm{V}$ of a test particle must be such that $|\bm{V}| \le 1$ at $\bm{X}=0$. 


\subsection{$\phi_{\hat \alpha \hat \beta}=0$}

An important special case involves a fiducial observer that has vanishing acceleration tensor. Indeed, in this case the reference path $\tilde{x}^\mu(\tau)$ is a timelike geodesic and the adapted orthonormal tetrad frame is parallel transported along the fiducial geodesic world line, i.e. $D\lambda^\mu{}_{\hat \alpha}/d\tau=0$.  Therefore, in Eqs.~\eqref{B6} and~\eqref{B7}, we have $P = 1$ and $\bm{Q} = 0$. Moreover, Eq.~\eqref{B14} can be written in this special case as
\begin{eqnarray}\label{B16} 
 \frac{d^2X^{\hat i}}{dT^2}&+&^FR_{\hat 0\hat i\hat 0\hat j}X^{\hat j}+2\,^FR_{\hat i\hat k\hat j\hat 0}V^{\hat k}X^{\hat j}\nonumber \\
 &+&\frac{2}{3}\,\left(3\,^FR_{\hat 0\hat k\hat j\hat 0}V^{\hat i}V^{\hat k}
+ \,^FR_{\hat i\hat k\hat j\hat l}V^{\hat k}V^{\hat l}+ \,^FR_{\hat 0\hat k\hat j\hat l}V^{\hat i}V^{\hat k}V^{\hat l}\right) X^{\hat j} +~ \mathcal{O}(|\mathbf{X}|^2) = 0\,.
\end{eqnarray}
This geodesic deviation equation is a generalized Jacobi equation~\cite{Chicone:2002kb}. Limited to a particular direction in space, say, $X^{\hat 1} = \Psi$ and $V^{\hat 1} = d\Psi/dT := \dot{\Psi}$, Eq.~\eqref{B16} reduces to 
\begin{equation}\label{B17}
\ddot{\Psi} + \,^FR_{\hat 0 \hat 1 \hat 0 \hat 1}\, (1- 2\,\dot{\Psi}^2) \Psi + \mathcal{O}(\Psi^2) = 0\,. 
\end{equation} 


\section{$\Lambda > 0$, $\mu = p = 0$}

For $\mu = p = 0$, the gravitational field equations reduce to
\begin{equation}\label{C1}
\frac{\ddot a}{a} = \frac{1}{3}\, \Lambda\,,
\end{equation}
\begin{equation}\label{C2}
\dot{a}^2 + k = \frac{1}{3}\, \Lambda\,a^2\,,
\end{equation}
where we assume the cosmological constant $\Lambda$ does not vanish. 

Let us first assume that $\Lambda > 0$. Then, with
\begin{equation}\label{C3}
(\frac{1}{3}\,\Lambda)^{\frac{1}{2}} = h > 0\,, \qquad a = a_{+}\, e^{ht} + a_{-}\, e^{-ht}\,,
\end{equation}
Eq.~\eqref{C2} implies
\begin{equation}\label{C4}
4 h^2 a_{+} a_{-} = k\,.
\end{equation}
Substituting this general solution in the FLRW metric~\eqref{I5} leads to de Sitter's spacetime. 

In view of the observed recession of distant galaxies, we can use a simple shortcut by assuming $a_{-} = 0$.  Hence, $k = 0$ and the resulting FLRW metric~\eqref{I5} with the constant scaling $a_{+} x^i \to x^i$ and $h \to H$ describes de Sitter's spacetime
\begin{equation}\label{C5}
ds_{\rm de\,Sitter}^2 = -dt^2 + e^{2Ht} \,\delta_{ij} dx^i dx^j\,.
\end{equation}

Returning to the general case, let us write Eq.~\eqref{C3} as
\begin{equation}\label{Ca}
a(t) = \hat{a} \,\cosh(ht + \hat{b})\,,
\end{equation}
where $\hat{a}$ and $\hat{b}$ are integration constants. Then, Eq.~\eqref{C2} implies $k = h^2 \hat{a}^2$. Employing metric~\eqref{I5} with $t \to t - \hat{b}/h$ and spherical coordinates $(r, \hat{\theta}, \hat{\phi})$, we get
\begin{equation}\label{Cb}
ds^2 = -dt^2 + \frac{\hat{a}^2\,\cosh^2(ht)}{(1 + \tfrac{1}{4}\,h^2 \hat{a}^2 r^2)^2}\,(dr^2 + r^2 d \hat{\Omega}^2)\,,
\end{equation}
where
\begin{equation}\label{Cc}
d \hat{\Omega}^2 = d\hat{\theta}^2 + \sin^2 \hat{\theta}\,d \hat{\phi}^2\,.
\end{equation}
Define
\begin{equation}\label{Cd}
\frac{1}{2}\, h\, \hat{a}\,r := \tan(\tfrac{1}{2} \chi)\,;
\end{equation}
then, metric~\eqref{Cb} takes the form
\begin{equation}\label{Ce}
ds^2 = -dt^2 + \frac{1}{h^2}\,\cosh^2(ht)\,(d \hat{\chi}^2 + \sin^2 \hat{\chi}\, d \hat{\Omega}^2)\,.
\end{equation}
Next, it is straightforward to verify that the coordinate transformation given on page 125 of Ref.~\cite{HE} connects metrics~\eqref{Ce} and~\eqref{C5}; that is, metric~\eqref{Ce} takes the form of de Sitter's metric~\eqref{C5} with $h \to H$.

\subsection{Exact Fermi Coordinates for de Sitter's Spacetime}

In de Sitter's spacetime with metric~\eqref{C5}, we consider a comoving observer with adapted orthonormal tetrad as in Eq.~\eqref{I13}. To simplify matters, we assume the observer is at the origin of spatial coordinates: $x^i = 0, i = 1, 2, 3$. The establishment of an exact Fermi normal coordinate system about this observer is described in detail in~\cite{Chicone:2005vn}. The exact coordinate transformation $(t, \bm{x}) \to (T, \bm{X})$ is given by~\cite{Chicone:2005vn}
\begin{equation}\label{C6}
e^{Ht} = e^{HT}\, \cos(H\mathbb{R})\,, \qquad \bm{x} = e^{-HT} \frac{\tan(H\mathbb{R})}{H\mathbb{R}}\,\bm{X}\,,
\end{equation}
where $\mathbb{R} = |\bm{X}|$. The Fermi coordinates are admissible for $0 \le \mathbb{R} < \pi/(2H)$. Introducing spherical Fermi coordinates via $X^{\hat1} = \mathbb{R} \,\sin\Theta_F \,\cos \Phi_F$, $X^{\hat 2} = \mathbb{R} \,\sin\Theta_F \,\sin \Phi_F$  and $X^{\hat 3} = \mathbb{R}\, \cos \Theta_F$, the Fermi metric for de Sitter spacetime takes the form~\cite{Chicone:2005vn}
\begin{equation}\label{C7}
^Fds^2_{\rm  de~Sitter} = - \cos^2(H\mathbb{R})\,dT^2 + d\mathbb{R}^2 + \frac{\sin^2(H\mathbb{R})}{H^2}\,d\Omega_F^2\,,
\end{equation}
where
\begin{equation}\label{C8}
d\Omega_F^2 = d\Theta_F^2 + \sin^2 \Theta_F \, d\Phi_F^2\,.
\end{equation}

Let us first note that using $\cos x \approx 1-\frac{1}{2} x^2$ and $\sin x \approx x - \frac{1}{6} x^3$, metric~\eqref{C7} can be reduced to metric~\eqref{I21} with $\mathbb{R} = \rho$ and $(\Theta_F, \Phi_F) = (\vartheta, \varphi)$. Next, let $T = \mathfrak{t}$ and define a new radial coordinate $\mathfrak{r}$ such that 
\begin{equation}\label{C9}
\mathfrak{r} := \frac{\sin(H\mathbb{R})}{H}\,.
\end{equation}
Then, metric~\eqref{C7} reduces to the standard form of de Sitter's metric, namely, 
\begin{equation}\label{C10}
^Fds_{\rm de\,Sitter}^2 = - (1- \frac{1}{3} \Lambda \mathfrak{r}^2) d\mathfrak{t}^2 + \frac{d\mathfrak{r}^2}{1- \frac{1}{3} \Lambda \mathfrak{r}^2} + \mathfrak{r}^2 \,d\Omega_F^2\,.
\end{equation}

For $\Lambda < 0$ in Eq.~\eqref{C10}, we get the standard form of the anti-de Sitter (AdS) spacetime; that is,  
\begin{equation}\label{C11}
^Fds_{\rm AdS}^2 = - (1+\frac{1}{3} |\Lambda| \mathfrak{r}^2) d\mathfrak{t}^2 + \frac{d\mathfrak{r}^2}{1+ \frac{1}{3} |\Lambda| \mathfrak{r}^2} + \mathfrak{r}^2 \,d\Omega_F^2\,.
\end{equation}
Let 
\begin{equation}\label{C12}
\left(\frac{1}{3} |\Lambda|\right)^{\frac{1}{2}}:= \varpi\,;
\end{equation}
then, with $\mathfrak{t} = T$ and 
\begin{equation}\label{C13}
\mathfrak{r} := \frac{\sinh(\varpi\mathbb{R})}{\varpi}\,,
\end{equation}
Eq.~\eqref{C11} reduces to the exact Fermi metric for the AdS spacetime~\cite{KC}
\begin{equation}\label{C14}
^Fds^2_{\rm  AdS} = - \cosh^2(\varpi\mathbb{R})\,dT^2 + d\mathbb{R}^2 + \frac{\sinh^2(\varpi\mathbb{R})}{\varpi^2}\,d\Omega_F^2\,.
\end{equation}
The construction of the exact Fermi coordinate system for the AdS spacetime is due to Klein and Collas~\cite{KC}. Appendix D contains a different derivation of the exact Fermi coordinates for the AdS spacetime along the lines of Ref.~\cite{Chicone:2005vn}.


\section{$\Lambda < 0$, $\mu = p = 0$}

In this case, the gravitational field equations~\eqref{C1} and~\eqref{C2} imply
\begin{equation}\label{D1}
a = a_{1}\, \sin(\varpi t) + a_{2}\, \cos(\varpi t)\,, \qquad \varpi^2 (a_1^2 + a_2^2) + k = 0\,. 
\end{equation}
Therefore, 
\begin{equation}\label{D2}
k = -1\,, \qquad a_1 = \frac{1}{\varpi}\,\cos\phi_0\,, \qquad a_2 = \frac{1}{\varpi} \, \sin \phi_0\,, 
\end{equation}
where $\phi_0$ is a constant angle.  With a simple temporal translation in the FLRW metric~\eqref{I5}, we can set $\phi_0 = 0$ such that $a(t) = \sin(\varpi t)/\varpi$. The resulting FLRW metric in this case describes the anti-de Sitter (AdS) spacetime
\begin{equation}\label{D3}
ds_{\rm AdS}^2 = -dt^2 + \frac{\sin^2(\varpi t)}{\varpi^2 (1-\frac{1}{4}\,r^2)^2}\,\delta_{ij} dx^i\,dx^j\,.
\end{equation} 

Regarding the range of validity of coordinates in the AdS metric~\eqref{D3}, we note, for instance, that they are admissible for $0 < t < \pi / \varpi$ and $0 \le r < 2$.


\subsection{Exact Fermi Coordinates for the AdS Spacetime}

We start with metric~\eqref{D3} and establish an exact Fermi normal coordinate system along the world line of a fiducial comoving observer with adapted tetrad~\eqref{I13}. As before, for the sake of simplicity, we assume the fiducial observer is at the origin of spatial coordinates: $x^i = 0, i = 1, 2, 3$. Next, we must work out the spacelike geodesics of metric~\eqref{D3}. 

Let $\sigma$ be the proper length of a spacelike geodesic. Then, the geodesic equations of motion are 
\begin{equation}\label{D4}
\frac{d^2t}{d\sigma^2} + a \dot{a} \frac{v^2}{f^2} = 0\,,
\end{equation}
\begin{equation}\label{D5}
\frac{d^2x^i}{d\sigma^2} + 2 \frac{\dot a}{a}\,\frac{dt}{d\sigma}\,v^i +\frac{1}{2f}[2(\bm{x} \cdot \bm{v}) v^i -v^2 x^i] = 0\,,
\end{equation}
where $v^i := dx^i/d\sigma$, $v^2 := \delta_{ij} v^i v^j$ and we have used the results of Appendix A. Moreover, the spacelike geodesic has unit length; hence, 
\begin{equation}\label{D6}
- \left(\frac{dt}{d\sigma}\right)^2 + a^2\frac{v^2}{f^2} = 1\,.
\end{equation}
Equations~\eqref{D4} and~\eqref{D6} together imply
\begin{equation}\label{D7}
\frac{1}{a^2} \,\left[1 + \left(\frac{dt}{d\sigma}\right)^2\right] = - \frac{1}{a \dot a} \,\left(\frac{dt}{d\sigma}\right)\frac{d}{dt}\left(\frac{dt}{d\sigma}\right) = \frac{v^2}{f^2}\,.
\end{equation}
The differential equation for $dt/d\sigma$ can be solved straightforwardly and the result is
\begin{equation}\label{D8}
\cos(\varpi t) = \alpha \,\cosh(\varpi \sigma + \ell)\,,
\end{equation}
where $\alpha$ and $\ell$ are constants of integration. 

As described in Appendix B, we are interested in a spacelike geodesic segment that starts at event $\mathbb{O}$ on the world line of the fiducial observer where $\sigma = 0$ and extends to length $\sigma$ at event $\mathbb{P}$. The spacetime coordinates at $\mathbb{O}$ are $(\check{t}, 0, 0, 0)$ and at $\mathbb{P}$ are $(t, x^1, x^2, x^3)$. The geodesic segment has a tangent vector $\xi^\mu$ at $\sigma = 0$ that is orthogonal to the world line of the fiducial observer; that is, 
\begin{equation}\label{D9}
\xi^\mu = (\frac{dt}{d\sigma}, \bm{v})_{\sigma = 0}\,, \qquad \xi_\mu \chi^{\mu}{}_{\hat 0} = 0\,. 
\end{equation}
This means $(dt/d\sigma)_{\sigma = 0} = 0$. It then follows from Eq.~\eqref{D8} that $\ell = 0$ and we have 
\begin{equation}\label{D10}
\cos(\varpi \check t) = \alpha\,, \qquad \frac{v^2}{f^2} = \frac{\varpi^2 (1-\alpha^2)}{[1- \alpha^2 \cosh^2(\varpi \sigma)]^2}\,.
\end{equation}

To solve the spatial part of the spacelike geodesic equation, we assume
\begin{equation}\label{D11}
x^i (\sigma) = \frac{C^i}{C}\,\mathcal{F}(\sigma)\,, \qquad v^i = \frac{C^i}{C}\,\frac{d\mathcal{F}}{d\sigma}\,,
\end{equation}
where $C^i, i = 1, 2, 3,$ are constants and 
\begin{equation}\label{D12}
C = (\delta_{ij} C^i C^j)^\frac{1}{2}\,.
\end{equation}
Therefore, 
\begin{equation}\label{D13}
r^2 = \delta_{ij} x^i x^j = \mathcal{F}^2\,, \qquad v^2 = \left(\frac{d\mathcal{F}}{d\sigma}\right)^2\,,
\end{equation}
\begin{equation}\label{D14}
\frac{v^2}{f^2} = \frac{\left(\frac{d\mathcal{F}}{d\sigma}\right)^2}{[1-\frac{1}{4} \mathcal{F}^2]^2} =  \frac{\varpi^2 (1-\alpha^2)}{[1- \alpha^2 \cosh^2(\varpi \sigma)]^2}\,.
\end{equation}
The differential equation above for $\mathcal{F}(\sigma)$ has the solution
\begin{equation}\label{D15}
\mathcal{F}(\sigma)  =  \frac{2 \Sigma}{ 1 + \sqrt{1- \Sigma^2}}\,, \qquad \Sigma := \frac{\tanh(\varpi \sigma)}{\sqrt{1-\alpha^2}}\,.
\end{equation}
With this result, Eq.~\eqref{D5} is satisfied as well. Moreover, for $\sigma = 0$, we have $\mathcal{F}= 0$ and this is consistent with the initial event of the spacelike segment, which has coordinates $(t, x^1, x^2, x^3)_{\mathbb{O}} = (\check{t}, 0, 0, 0)$, where $\cos(\varpi \check t) = \alpha$. The endpoint of the spacelike segment at $\mathbb{P}$ has spacetime coordinates $(t, x^i)$ and Fermi coordinates $(T, X^{\hat i})$, where $T = \check{t}$ and $X^{\hat i} = \sigma \,\xi^\mu \chi_{\mu}{}^{\hat i}$. 
From
\begin{equation}\label{D16}
\frac{d\mathcal{F}}{d\Sigma}  =  \frac{2}{ (1 + \sqrt{1- \Sigma^2}) \sqrt{1-\Sigma^2}}\,, \qquad \frac{d\Sigma}{d\sigma} = \frac{\varpi}{\sqrt{1-\alpha^2}\,\cosh^2(\varpi \sigma)}\,,
\end{equation}
we find $(d\mathcal{F}/d\sigma)_{\sigma = 0} = \varpi/\sqrt{1-\alpha^2}$ and hence $X^{\hat i} = \sigma\,C^i/C$. The transformation between spacetime and Fermi coordinates is thus given by
\begin{equation}\label{D17}
\cos(\varpi t) = \cos(\varpi \check{t}) \,\cosh(\varpi \sigma)\,, \qquad  x^i =  \frac{\mathcal{F}}{\sigma} X^{\hat i}\,.
\end{equation}

To write the AdS metric~\eqref{D3} in terms of Fermi coordinates, let us first note that $\check{t} = T$ and 
\begin{equation}\label{D18}
\sigma = (\delta_{ij} X^{\hat i} X^{\hat j})^\frac{1}{2} = \mathbb{R}\,
\end{equation}
is the radial Fermi coordinate. Therefore, from Eq.~\eqref{D17} we have 
\begin{equation}\label{D19}
\cos(\varpi t) = \cos(\varpi T) \,\cosh(\varpi \mathbb{R})\,,
\end{equation}
which upon differentiation implies
\begin{equation}\label{D20}
dt = \frac{1}{\sin(\varpi t)}\,[\sin(\varpi T) \,\cosh(\varpi \mathbb{R}) dT - \cos(\varpi T) \,\sinh(\varpi \mathbb{R}) d\mathbb{R}]\,.
\end{equation}
Next, in the solution for spacelike geodesics, $\mathcal{F}$ was only a function of $\sigma = \mathbb{R}$ and $\alpha = \cos(\varpi T)$ was treated as a constant; however, in the present context, $\mathcal{F}$ is a function of both $\mathbb{R}$ and $T$. From Eq.~\eqref{D17}, we find
\begin{equation}\label{D21}
dx^i = d\left(\frac{\mathcal{F}}{\mathbb{R}}\right)\,X^{\hat i} + \frac{\mathcal{F}}{\mathbb{R}}\,dX^{\hat i}\,.
\end{equation}
Let us introduce spherical Fermi coordinates such that 
\begin{equation}\label{D22}
\delta_{ij} dX^{\hat i}\, dX^{\hat j} = d\mathbb{R}^2 + \mathbb{R}^2 \,d\Omega_F^2\,;
\end{equation}
then, after some algebra we get
\begin{equation}\label{D23}
\delta_{ij} dx^i\,dx^j = (\mathcal{F}_T dT + \mathcal{F}_{\mathbb{R}} d \mathbb{R})^2 + \mathcal{F}^2\,d\Omega_F^2\,,
\end{equation} 
where $\mathcal{F}_T = \partial \mathcal{F}/ \partial T$, etc. 
From 
\begin{equation}\label{D24}
f  =  \frac{2 \sqrt{1- \Sigma^2}}{ 1 + \sqrt{1- \Sigma^2}}\,, \qquad \frac{a^2 \mathcal{F}^2}{f^2} = \frac{\sinh^2(\varpi \mathbb{R})}{\varpi^2}\,,
\end{equation}
\begin{equation}\label{D25}
\frac{\mathcal{F}_{\mathbb{R}}}{f}  =  \frac{\varpi \sin(\varpi T)}{\sin^2(\varpi t)}\,, \qquad \frac{\mathcal{F}_T}{f} = -\frac{\varpi \cos(\varpi T)}{\sin^2(\varpi t)}\,\sinh(\varpi \mathbb{R}) \,\cosh(\varpi \mathbb{R})\,,
\end{equation}
we can calculate the spatial part of the AdS metric~\eqref{D3} in terms of Fermi coordinates, namely,
\begin{align}\label{D26}
\frac{\sin^2(\varpi t)}{\varpi^2 f^2}\,\delta_{ij} dx^i\,dx^j = &{} \frac{[\cos(\varpi T) \,\sinh(\varpi \mathbb{R})\,\cosh(\varpi \mathbb{R})\,dT- \sin(\varpi T)\,d\mathbb{R}]^2}{\sin^2(\varpi t)} \nonumber  \\
&{} + \frac{\sinh^2(\varpi \mathbb{R})}{\varpi^2 }\,d\Omega_F^2\,.
\end{align} 
Finally, we can write the AdS metric in Fermi coordinates by combining the above result with Eq.~\eqref{D20}. Using the relation
\begin{equation}\label{D27}
\sin^2(\varpi t) = \sin^2(\varpi T) - \cos^2(\varpi T)\, \sinh^2(\varpi \mathbb{R})\,,
\end{equation} 
 we find, as in Appendix C,   
\begin{equation}\label{D28}
^Fds^2_{AdS} = -dt^2 + \frac{\sin^2(\varpi t)}{\varpi^2 f^2}\,\delta_{ij} dx^i\,dx^j = - \cosh^2(\varpi \mathbb{R})\,dT^2 + d\mathbb{R}^2 + \frac{\sinh^2(\varpi \mathbb{R})}{\varpi^2 }\,d\Omega_F^2\,.
\end{equation}

\end{document}